# Designing Mixed-Metal Electrocatalyst Systems for Photoelectrochemical Dinitrogen Activation

Manpreet Kaur[1], Marc Walker[2], Steven Hindmarsh[3], Charlotte Bolt[1], Stephen York[3], Yisong Han[3], Martin R. Lees[2], Katharina Brinkert*[1,4,z]

[1]Department of Chemistry, University of Warwick, Gibbet Hill Road, Coventry, CV4 7AL, United Kingdom
[2]Department of Physics, University of Warwick, Gibbet Hill Road, Coventry, CV4 7AL, United Kingdom
[3]Electron Microscopy Research Technology Platform, University of Warwick, Gibbet Hill Road, Coventry, CV4 7AL, United Kingdom
[4]Center for Applied Space Technology and Microgravity (ZARM), University of Bremen, 28359 Bremen, Germany

[z]katharina.brinkert@warwick.ac.uk

**Abstract**

Efficient artificial photosynthesis systems are currently realized as catalyst- and surface-functionalized photovoltaic tandem- and triple-junction devices, enabling photoelectrochemical (PEC) water oxidation while simultaneously recycling $CO_2$ and generating hydrogen as a solar fuel for storable renewable energy. Although PEC systems also bear advantages for the activation of dinitrogen - such as a high system tunability with respect to the electrocatalyst integration and a directly controllable electron flux to the anchoring catalyst through the adjustability of incoming irradiation - only a few PEC devices have been developed and investigated for this purpose.

We have developed a series of photoelectrodeposition procedures to deposit mixed-metal electrocatalyst nanostructures directly on the semiconductor surface for light-assisted dinitrogen activation. These electrocatalyst compositions containing Co, Mo and Ru in different atomic ratios follow previously made recommendations of metal compositions for dinitrogen reduction and exhibit different physical properties. XPS studies of the photoelectrode surfaces reveal that our electrocatalyst films are to a large degree nitrogen-free after their fabrication, which is generally difficult to achieve with traditional magnetron sputtering or e-beam evaporation techniques. Initial chronoamperometric measurements of the p-InP photoelectrode coated with the Co-Mo alloy electrocatalyst show higher photocurrent densities in the presence of $N_2(g)$ than in the presence of Ar at -0.09 V *vs* RHE. Indications of successful dinitrogen activation have also been found in consecutive XPS studies, where both, N 1s and Mo 3d spectra, reveal evidence of nitrogen-metal interactions.

# 1. Introduction

Photoelectrochemical (PEC) devices have emerged as stable and efficient systems to use for the light-assisted production of chemicals such as carbon-based fuels and hydrogen[1] that allow driving these reactions only by harnessing and converting solar energy. The devices are based on technically modified, integrated semiconductor-electrocatalyst systems which can be altered and optimised with respect to the thermodynamic and kinetic requirements of the desired chemical reaction. Given the high tunability and the electron flux control at the electrode-electrolyte interface through the incident light, PEC devices could potentially also provide an advantage in realising the electrochemical nitrogen reduction reaction (NRR) at high Faradaic efficiencies and current densities.[2] Generally, the electrochemical NRR bears several challenges such as the stable N≡N bond (942 kJ mol$^{-1}$) and a low product selectivity due to the competing, kinetically favoured hydrogen evolution reaction (HER). So far, only very few approaches report the successful electrochemical activation of dinitrogen ($N_2(g)$) and its reduction as well as consecutive protonation[3–11] as all suffer from low yields and reaction efficiencies. Crucial for realising the efficient and stable electrochemical NRR is the (photo-) electrode design and the choice of electrocatalyst material. Here, it is particularly important to avoid nitrogen contamination in the electrode manufacturing process as it poses challenges to the identification of effective catalyst materials for N≡N bond splitting. Ever since the identification of the bimetallic nature of Nitrogenases' catalytically active site (the most prominent one using an FeMo cofactor[12,13]) and the in-depth analysis of the catalytic activity dependency on the nitrogen adsorption energy through the Brønsted-Evans-Polanyi relationship, bimetallic electrocatalysts are discussed and investigated for the electrochemical NRR. After the initial suggestion of Jacobsen et al., the electrocatalyst combination of Co and Mo has been explored in particular for the NRR as the metal combination best balances the rate-limiting step of $N_2(g)$ dissociation and the stabilisation of adsorbed nitrogen on the electrode surface.[14–22] As an example, Tsuji et al. report Co-Mo alloy nanoparticles supported on $CeO_2$ which exhibit a higher catalytic activity for the NRR than the individual metals.[17] Li et al. also report on the synthesis of atomically dispersed bimetallic pairs of Co and Mo on nitrogen-doped carbon frameworks and state a promising NRR performance in an alkaline electrolyte with a Faradaic efficiency of 23.18%.[23] Crucial for the interpolated properties of these mixed-metal electrocatalyst systems is the close proximity of the metals on the electrode surface. This is usually not straight forward to realise as one of the components can segregate to the surface as a consequence of the properties of a binary alloy[14] or deposition techniques might not allow the precise control over the atomic arrangement and desired alloy formation.

Here, we report on the fabrication of mixed-metal Co-Mo and Co-Mo-Ru electrocatalyst systems directly on a p-type indium phosphide (p-InP) surface via photoelectrodeposition, opening up the possibility of the photoelectrochemical NRR without nitrogen contamination originating from the electrode fabrication step. By using different co- and consecutive photoelectrodeposition procedures as well as electrolyte additives such as boric acid and 2-propanol, we can control the local metal composition and surface morphology as well as the inherent magnetic properties of our photoelectrodes, leading to precisely tuneable mixed-metal systems for nitrogen reduction catalysis. Initial chronoamperometric measurements of selected, fabricated electrodes reveal higher photocurrent densities in the presence of $N_2(g)$ than Ar at - 0.09 V *vs* RHE. XPS studies provide furthermore first evidence of metal-nitrogen species on the surface, strongly indicating the activation of $N_2$(g).

## 2. Experimental

### 2.1 Chemicals
All chemicals were used without further modifications, including molybdenum (III) chloride ($MoCl_3$, 99.5%, Alfa Aesar), sodium molybdate dihydrate ($Na_2MoO_4 \cdot 2H_2O$, >99.5%, Fisher Scientific), cobalt (II) chloride, ($CoCl_2$, 99.7%, Alfa Aesar), cobalt (II) sulfate heptahydrate ($CoSO_4 \cdot 6H_2O$, 99%, Fisher Scientific), ruthenium (III) chloride hydrate ($RuCl_3 \cdot xH_2O$, >99.9%, Sigma Aldrich), boric acid ($H_3BO_3$, >99.5%, Sigma Aldrich), sodium chloride (NaCl, 99.5% +, VWR), and sodium sulfate ($Na_2SO_4$, >98.5%, VWR).

### 2.1 Preparation of the p-InP photoelectrodes
Single crystal p-InP wafers with the orientation (111A) were obtained from AXT Inc. (Geo Semiconductor Ltd. Switzerland) with a Zn doping concentration of $5 \times 10^{17}$ cm$^{-3}$. The preparation of an ohmic back contact involved the sputtering of 4 nm Au, 80 nm Zn and 150 nm Au on the backside of the wafer which was then heat-treated to 400 °C for 60 s. The 0.5 cm$^2$ polished indium face of p-InP was furthermore etched for 30 s in bromine (0.05% (w/v))/ methanol solution, rinsed with ethanol and ultrapure water and dried under argon flux. All solutions were made from ultrapure water and analytical grade chemicals. Subsequent cyclic voltammetric measurements were performed in a standard three-electrode potentiostatic arrangement where a Pt coil was used as counter and an Ag/AgCl (3 M KCl) (E = -0.210 V *vs* RHE) as a reference electrode. All potentials in the text are converted to those *vs* the Reversible Hydrogen Electrode (RHE). All electrolytes were purged with Ar of research grade purity for 15 mins prior to usage. The p-InP surface was photoelectrochemically conditioned in 0.5 M HCl(aq) by potentiodynamic cycling under illumination (100 mW/cm$^2$) between -0.44 V and +0.31 V at a scan rate of 50 mV/s while purging with Ar of research grade purity. Illumination occurred with a Xe arc lamp (Edmund Optics) through a quartz window of the borosilicate glass cell. The light intensity was adjusted with a calibrated silicon reference photodiode (Hamamatsu S1227-66BQ) to 100 mW cm$^{-2}$.

### 2.2 Photoelectrodeposition of Co, Mo and Ru
Chronoamperometric (CA) methods were employed for metal photoelectrodeposition using a BioLogic SP-200 potentiostat. In a three-electrode system, the etched and preconditioned p-InP, Pt coil, and Ag/AgCl (3 M KCl) electrodes were used as the working electrode, counter electrode and reference electrode, respectively. Compositions of utilised deposition electrolytes are described in **Table 1**. All solutions were prepared using MilliQ water (18.2 MΩ) and purged for 15 min with research grade Ar before utilisation. The photoelectrodeposition occurred under constant magnetic stirring and Ar purging at the indicated potentials and times in **Table 1**. The deposition potential was applied from the open-circuit potential of the electrode. Illumination occurred with a Xe arc lamp (Edmund Optics) through a quartz window of the borosilicate glass cell at 100 mW cm$^{-2}$. Each electrode was prepared at least three times to ensure reproducibility. To avoid chemisorption processes, the working electrode was immersed into the solution just a few ms before potential application. All fabricated electrodes do not show indications of nitrogen on the surface. Sample N 1s spectra (overlapping with the Mo 3p spectrum) are shown in **Fig. SI 1** in the Supporting Information (SI).

| Method of Co/Mo deposition | Metal electrolyte | Electrolyte composition | pH | Time |
|---|---|---|---|---|
| *Consecutive depositions* | Mo solution I | $Na_2MoO_4 \cdot 2H_2O$ (10 mM)<br>$Na_2SO_4$ (1 M)<br>$H_3BO_3$ (0.5 M) | 5.2 | 10 s |
| | Mo solution II | $MoCl_3$ (1 mM)<br>NaCl (0.5 M)<br>2-propanol (0.5% (v/v)) | 4 | 10 s |
| | Co solution I | $CoSO_4$ (0.2 M)<br>$Na_2SO_4$ (1 M)<br>$H_3BO_3$ (0.5 M) | 5.2 | 3 s |
| | Co solution II | $CoCl_2$ (0.1 M)<br>NaCl (0.5 M)<br>2-propanol (0.5% (v/v)) | 7.2 | 3 s |
| | Ru solution I | $RuCl_3$ (0.33 mM)<br>NaCl (0.5 M)<br>2-propanol (0.5% (v/v)) | 3.8 | 5 s |
| *Co-depositions* | Co-Mo solution I | $CoSO_4$ (0.2 M)<br>$Na_2MoO_4 \cdot 2H_2O$ (10 mM)<br>$Na_2SO_4$ (1 M)<br>$H_3BO_3$ (0.5 M) | 5.2 | 10 s |
| | Co-Mo solution II | $CoSO_4$ (0.2 M)<br>$Na_2MoO_4 \cdot 2H_2O$ (10 mM)<br>$Na_2SO_4$ (1 M) | 6.8 | 10 s |
| | Co-Mo-Ru solution | $CoCl_2$ (2 mM)<br>$MoCl_3$ (0.1 mM)<br>$RuCl_3 \cdot xH_2O$ (0.1 mM)<br>$H_3BO_3$ (0.1 M)<br>NaCl (0.5 M) | 3.5 | 5 s |

**Table 1.** Compositions of electrolytes used for the Co-Mo-Ru consecutive and co-photoelectrodepositions. The applied potential was set to -0.09 V *vs* RHE in all cases.

### 2.2.1 Consecutive photoelectrodepositions

Bimetallic depositions of Mo and Co on the p-InP photoelectrode were achieved by consecutively depositing the metals Mo and Co from the deposition electrolytes at the indicated concentrations, pH values and time durations specified in **Table 1**. Two deposition solutions, $Na_2MoO_4$(aq) (Mo solution I) and $MoCl_3$(aq) (Mo solution II) were investigated. For the consecutive Co deposition, two solutions were investigated: $CoSO_4$(aq) (Co solution I) and $CoCl_2$(aq) (Co solution II), respectively. By combining the different Mo and Co solutions, photoelectrodes with four different coatings were prepared:

1. $Na_2MoO_4$ -$CoCl_2$ (deposition from Mo solution I followed by Co solution II);
2. $Na_2MoO_4$ -$CoSO_4$ (deposition from Mo solution I followed by Co solution I);
3. $MoCl_3$ -$CoSO_4$ (deposition from Mo solution II followed by Co solution I); and
4. $MoCl_3$ - $CoCl_2$ (deposition from Mo solution II followed by Co solution II).

### 2.2.2 Co-photoelectrodeposition of Co, Mo

For the co-photoelectrodepositions, Mo and Co were deposited onto p-InP out of the same electrolyte solution using Co-Mo solutions I and Co-Mo solution II (**Table 1**). This procedure solely involved the use of $Na_2MoO_4$(aq) and $CoSO_4$(aq) and the addition of boric acid in the case of Co-Mo solution I.

### 2.2.3 Co-photoelectrodeposition of Co, Mo and Ru

For the deposition of Co, Mo and Ru, two approaches were used: p-InP-CoMo electrodes were firstly prepared using the Co-Mo solution I described in **Table 1**. Subsequently, Ru was deposited on top of the p-InP-CoMo photoelectrodes using Ru solution I at the specified potential and time duration. p-InP-CoMoRu electrodes were also prepared by co-photoelectrodeposition of all three metals on the p-InP substrate using the Co-Mo-Ru electrolyte and deposition conditions specified in **Table 1**.

### 2.3 Photoelectrochemical dinitrogen activation in aqueous electrolyte

Initial photoelectrochemical CA measurements were performed using a BioLogic SP-200 potentiostat controlled by a standard EC-Lab software. The experiments were performed in a standard three-electrode potentiostatic arrangement in a sealable electrochemical cell. A Pt coil (ALS Co., Ltd) was used as the counter electrode and Ag/AgCl (3 M KCl, WPI Europe, DRIREF-5) was used as the reference electrode. Pt deposits were not observed on the electrode surface by XPS (working electrode) after electrochemical testing. The electrolyte was 50 mL of 50 mM aqueous $H_2SO_4$. Prior to the measurements, the electrolyte was purged for 15 min with Ar (5.0 purity) under gentle stirring to remove $O_2$(g) and $N_2$(g). To saturate the electrolyte with $N_2$ prior to electrochemical measurements, the electrolyte was purged for 15 min with 99.999% pure $N_2$. Before entering the electrochemical (EC) cell, $N_2$ and Ar were directed through a gas bubbler that contained 2.4 M $H_2SO_4$(aq). All photoelectrochemical tests were carried out for 20 min under constant $N_2$ (99.999%) purging with illumination provided by a Xe arc lamp (Edmund Optics) through a quartz window of the borosilicate glass cell at 100 mW $cm^{-2}$. Three freshly prepared electrodes were tested in total to ensure repetition.

### 2.4 Photoelectrode surface characterization

*2.4.1 Scanning electron microscopy (SEM)*
SEM images were obtained with a Zeiss SUPRA 55-VP instrument at varying accelerating voltages between 2kV and 10 kV and an in-lens secondary e-detector. The system had an Oxford Instruments energy-dispersive X-ray (EDX) spectrometer that was used for the elemental composition analysis. Each electrode was scanned at least at three different locations. Scans were conducted via point measurements as well as over a given area.

*2.4.2 Atomic force microscopy (AFM)*
AFM images were obtained on a Bruker Dimension Icon or a Multimode 8 instrument using Bruker ScanAsyst-Air probes (silicon tip, silicon nitride cantilever, spring constant: 0.4 N/m, frequency: 45-95 kHz), operating in the peakforce tapping mode. Images were analyzed using the *Gwyddion* software. Height distribution data were acquired from 2 μm x 2 μm AFM images masked using the watershed method. The masked images are included in the SI for reference (**Fig. S2**).

*2.4.3 X-ray photoelectron spectroscopy (XPS)*
XPS was performed using a Kratos Analytical Axis Ultra DLD spectrometer using a monochromatic Al Kα X-ray source (1486.69 eV) in a chamber with a base pressure below $1 \cdot 10^{-10}$ mbar. Samples were mounted on the sample bar using electrically conductive carbon tape. Survey spectra were collected using an analyzer pass energy of 160 eV at 1 eV increments, each of which were integrated for 200 ms. High resolution core level spectra were collected using a pass energy of 20 eV (resolution of approximately 0.4 eV) using multiple sweeps. Data were analysed using the CasaXPS package, employing mixed Gaussian-Lorentzian (Voigt) line shapes and Shirley backgrounds. The transmission function and work function of the spectrometer were calibrated using clean polycrystalline Ag foil. Binding energies for the different chemical species are given in **Table SI 1**.

*2.4.4 Transmission electron microscopy (TEM)*
Cross-sectional samples were prepared for TEM imaging using a Tescan Amber dual beam FIB (focused ion beam)-SEM. This instrument combines a FEG (field emission gun) SEM with a $Ga^+$ ion beam, both capable of 1 - 30 keV accelerating voltages. Electron beam deposition of a sacrificial Pt layer was performed, thus avoiding alteration of the important surface region by ion beam damage. A typical lift-out procedure was followed[24], with a small section of material milled out from the sample surface and mounted onto a TEM grid. Thinning of this lamella was performed, creating a ~4 μm wide electron transparent (i.e. under 100 nm thickness) region. Final polishing was undertaken at 5 keV, aiming to create a clean and suitably damage free sample. All images shown were taken using a JEOL 2100 TEM operated at 200 keV.

*2.4.5 Magnetometry*
Measurements of dc magnetisation, *M*, *vs* temperature, *T*, were conducted between 5 and 300 K and a heating, resp. cooling rate, of 2 K min$^{-1}$ in an applied field, *H*, of 250 Oe using a Quantum Design MPMS-5S superconducting quantum interference device (SQUID) magnetometer. Magnetisation *vs* field measurements were made between 5 and -5 kOe at 5 and 300 K. The films were mounted in plastic straws, with the plane of each film either parallel or perpendicular to the applied magnetic field. The SQUID magnetometer was calibrated using a standard palladium sample.

The density of the bimetallic deposits was assumed to be 10 g/cm$^3$ as it approximately represents the density of the individual Co and Mo metals. The molecular weight of the films was calculated using quantitative data from SEM-EDX analysis. The thickness of the films, to calculate the overall volume of the deposits, was established from TEM images, using the *ImageJ* software. The magnetic response of a bare p-InP substrate was measured and found to be diamagnetic (see **Figure S3**). These data were used to estimate the contribution to the magnetisation arising from the substrate when measuring the films.

## 3. Results and discussion

Chemical methods for metal deposition include electrodeposition[25,26], electroless deposition[27], photodeposition, and photoelectrodeposition. Several studies reporting Co and Mo (electro-) depositions can be found in the literature.[26,28–35] Lin et al. reported on achieving highly compact and smooth CoMo alloys via electrodeposition on carbon fibre paper (CFP).[35] The deposition potential was varied between -1.0 V to -1.4 V *vs* RHE at an interval of 0.2 V for 1200 s. XPS characterization showed Co and Mo present in several oxidation states, including Co$^0$ and Mo$^0$. Non-metallic states were attributed to be a result of exposure to ambient environments. Furthermore, Milikić et al. report on a 1 μm CoMo film obtained on a copper substrate using a 30 min electroless deposition method that yielded a current density of 10 mA cm$^{-2}$ for HER at an overpotential of 61 mV *vs* RHE. [27] Photoelectrodeposition protocols for cobalt exist,[36,37] as well as molybdenum disulfide,[38] however not for elementary molybdenum. Souche et al. reported on photoelectrodeposition of patterned Co films on p-Si substrates via cyclic voltammetry (CV) and pulsed deposition at constant potential of -0.859 V *vs* RHE and a pulse width of 100 to 200 ms, with a resting potential held below the deposition potential.[36] The illumination source in this study was a krypton laser with a power density of 60 mW cm$^{-2}$ (647 nm).

Generally, photo-assisted electrodepositions remain a vastly underexplored field with little literature,[36,39–42] and to the best of our knowledge, no reports of mixed-metal photoelectrodepositions for electrocatalyst syntheses exist. For our studies we decided to use p-InP as a substrate semiconductor as it has an optimal bandgap and band-edge positions for the reduction of dinitrogen to ammonia (+0.092 V *vs* RHE[19]): the non-doped InP has a band gap of 1.35 eV, with the valence band and conduction band edges positioned at -0.25 V and 1.10 V *vs* RHE, respectively.[43] Furthermore, the theoretically achievable maximum photocurrent density of 35 mA cm$^{-2}$ (AM 1.5 G) allows for sufficiently high current densities in future NRR tests. We employed both, consecutive and co-photoelectrodepositions of Co, Mo and Ru directly on the etched and electrochemically conditioned p-InP surface (see Experimental Section) in order to explore the achievable element ratios, their accompanying morphologies as well as the conditions leading to the deposition of mixed-metal and alloy electrocatalyst layers. A set of four p-InP-Mo-Co electrodes was prepared using consecutive depositions (Section 3.1) and a set of two p-InP-CoMo electrodes were prepared using co-deposition protocols (Section 3.2). Two Mo-based electrolytes were explored for Mo deposition, Na$_2$MoO$_4$(aq) and MoCl$_3$(aq), whereas CoSO$_4$(aq) and CoCl$_2$(aq)-based electrolytes were explored for Co deposition. For the deposition of Co, Mo and Ru (Section 3.3), two approaches were chosen: firstly, Mo and Co were co-photoelectrodeposited onto the p-InP surface and Ru was deposited afterwards using a $RuCl_3$(aq) solution and secondly, all three metals were deposited out of the same electrolyte solution containing $CoCl_2$(aq), $MoCl_3$(aq) and $RuCl_3$(aq).

The precise compositions of the deposition electrolytes are provided in the Experimental Section (**Table 1**). Due to the light-assisted nature of the deposition processes, the Co-, Mo- and Ru-ion concentrations in the deposition electrolytes had to be optimised with respect to solution transparency and light penetration, ensuring that the electrode surface was fully illuminated at a light intensity of 100 mW/cm$^2$ during the photoelectrodeposition process. The literature value for $Co^{2+}$ to Co reduction is -0.277 V *vs* RHE.[25] Reduction potentials for $Co^{2+}$(aq) at 0.1 M, for $CoCl_2$, and 0.2 M, for $CoSO_4$, were calculated using the Nernst equation and were found to be -0.247 V *vs* RHE, and -0.256 V *vs* RHE, respectively. It should be noted that these reduction potentials do not consider the protonation of $Co^{2+}$(aq) in acidic solutions. It was found that both deposition solutions (Mo solution I and Mo solution II) exclusively contained Mo (VI) species which have a reduction potential for $Mo^0$ formation of +0.075 V *vs* RHE.[44] The determination of the exact reduction potentials of the Mo solutions used here is difficult due to the complex Mo solution chemistry and the presence of multiple complex species in the electrolyte as illustrated below. The deposition process occurs through electron transfer from the semiconductor conduction band (and/or via surface states) to the metal ions in solution, whereas the utilization of light increases the electron density in the conduction band.[45–47] Therefore, the applied potential of -0.09 V *vs* RHE and the generated p-InP photovoltage provide a sufficient overpotential for all deposition reactions. It has to be noted that during all depositions, the HER was a competing process and $H_2$(g) evolution took place, influencing the transient photocurrent-time characteristics (*J-t*).

## 3.1 Consecutive photoelectrodepositions of Mo and Co

Historically, the electrodeposition of metallic molybdenum is known to be difficult due to its complex chemistry in aqueous solutions and its affinity to oxygen.[48,49] In weakly acidic solutions such as present in Mo solution I (pH 5.2), the tetrahedral molybdate anion $MoO_4^{2-}$ polymerizes and forms heptamolybdates $[Mo_7O_{24}]^{6-}$:[50,51]

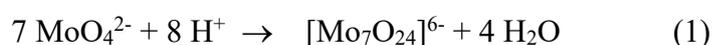

$$7\ MoO_4^{2-} + 8\ H^+ \rightarrow\ [Mo_7O_{24}]^{6-} + 4\ H_2O \qquad (1)$$

During electrodeposition without further electrolyte additives, the Mo (VI) is reduced to lower oxidation states, resulting in the formation of (hydrated) oxide layers on the electrode surface. Mo has a generally low hydrogen overpotential (+0.303 V), resulting in a major part of the applied voltage being used for the discharge of protons to form hydrogen during the HER. The in turn increased formation of $OH^-$ ions in proximity of the electrode surface can be removed by the addition of boric acid, shifting the equilibrium of the acid dissociation steps towards the formation of $H_3BO_3$.[52] Thus, Mo oxide and hydroxide formation can largely be prevented during the deposition reaction, favouring $Mo^0$ deposition. The boric acid also fulfils a second role: excess, undissociated $H_3BO_3$ has been hypothesised to directly interact with the electrode surface through adsorption and thereby blocking the growth of metal nuclei.[53,54] This leads to an increased nucleation rate during the deposition process and an overall smoother film with smaller, more homogenously shaped nanoparticles. $MoCl_3$(aq) is not water soluble, but the suspension quickly oxidises and forms blue polyoxometalates containing Mo (VI) as shown in XPS studies (**Fig. SI 4**).[55] **Fig. 1** (a) shows the different chronoamperometric characteristics of Mo photoelectrodepositions from both, $Na_2MoO_4$(aq) and $MoCl_3$(aq) solutions, as part of the consecutive depositions. Whereas both solutions give rise to a similar maximum photocurrent density ($J$) response (2 mA cm$^{-2}$), the transient photocurrent density characteristics for the $Na_2MoO_4$(aq) solution shows a slow decay and the $J$-$t$ curve for the $MoCl_3$(aq) solution plateaus after the initial photocurrent density decrease at about 0.5 s. In both cases, the initial peak represents the charge/discharge processes within the double layer and capacitance changes of the space-charge layer of the semiconductor, followed by the nucleation and growth current. Consecutive Co depositions result in significantly higher photocurrent densities (**Fig. 1** (b) and (c)) and very different transient $J$-$t$ behaviours despite similar $Co^{2+}$ reduction potentials: whereas for both p-InP- Mo surfaces, the Co deposited from a $CoSO_4$(aq) solution leads to photocurrent densities up to about 10 mA cm$^{-2}$, the photocurrent densities achieved with Co deposited from $CoCl_2$(aq) show significantly lower maximum values of up to 5 - 6 mA cm$^{-2}$ and a different $J$-$t$ behaviour. This illustrates the importance of electrolyte additives during the deposition process: the impact of $Cl^-$ ions on the deposition of Cu has been widely studied and it has been shown that a high concentration of $Cl^-$(aq) in the deposition electrolyte ($\geq$ 0.1 M) can cause the complexation of $Cu^{2+}$ ions in solution, leading to a net concentration decrease of free metal ions in solution which in turn induces a cathodic polarization of the deposition process.[56] Although this effect is not investigated for $Co^{2+}$(aq) solutions, it is hypothesised that the high $Cl^-$(aq) concentration present in the $CoCl_2$(aq) electrolyte here (Co solution II) causes a similar effect. This could explain the significantly lower photocurrent densities observed in the Co deposition process when Co solution II was used. The initially sharp, broad peak (0.5 - 0.75 s) during Co deposition from $CoSO_4$(aq) onto the p-InP - Mo surface also indicates a larger double-layer capacitance, which is independent of the Mo layer deposited prior to the Co deposition step.

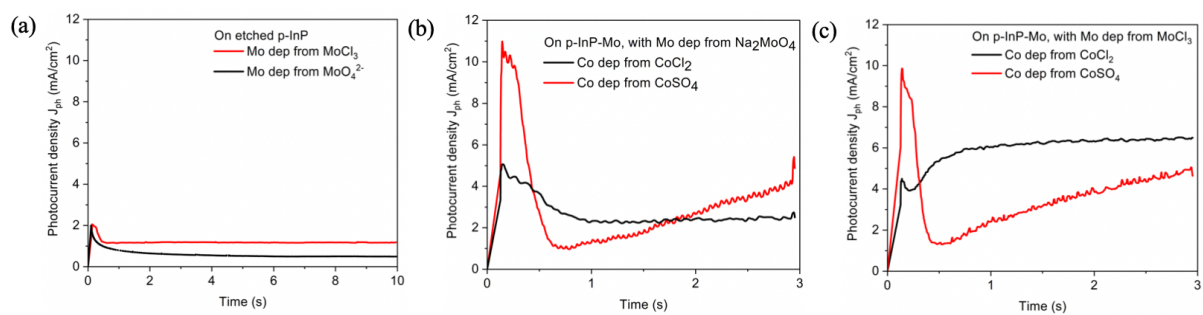

**Fig. 1** Chronoamperometric (CA) plots for the consecutive photoelectrodepositions of Mo and Co onto the modified p-InP surface from electrolytes specified in **Table 1**. Mo was firstly deposited from $MoCl_3$(aq) and $MoO_4^{2-}$(aq) solutions (a) followed by Co deposition onto both surfaces from $CoCl_2$(aq) and $CoSO_4$(aq) solutions, respectively, shown in (b) and (c). All $J(t)$ data were recorded at a potential of -0.09 V *vs* RHE. Illumination occurred during the depositions via an Xe arc lamp at a power density of 100 mW/cm$^2$.

### 3.1.1 Consecutive photoelectrodepositions from Na$_2$MoO$_4$ - CoCl$_2$ electrolytes

**Figure 2** shows the optical analyses of the photoelectrodes prepared by consecutive Mo - Co photoelectrodepositions from the Na$_2$MoO$_4$ and CoCl$_2$ electrolytes (Mo solution I, Co solution II). XPS studies of the smooth Mo film (particle sizes are about 18 - 22 nm) reveal the Mo oxidation states Mo$^{4+}$ (62 %) and Mo$^{6+}$ (33%, **Fig. S5** (a) and (b)) at 230.6 eV/233.7 eV and 232.0 eV/235.1 eV, respectively. As confirmed by EDX analysis, the large grains on top of the film are Co deposits with the oxidation states Co$^0$ (5%, 777.5 eV) and Co$^{2+}$ (95%). The two peaks at 780.9 eV/ 782.7 eV and 786.5 eV/ 790.9 eV in the XP spectrum (**Fig. S5** (c)) reveal that Co$^{2+}$ is present on the surface as a hydroxide. Mo$^0$ is only found to be present on the surface with 5% (229.1 eV/232.2 eV), which is generally in agreement with existing literature stating the difficulty of electrochemically depositing Mo$^0$. The overall Co:Mo at% ratio on the electrode surface has been found to be 70:30 in EDX analysis. It is evident that this consecutive deposition protocol results in a clear separation of Co and Mo on the electrode surface and the deposition of Mo and Co in various metal oxidation states. The height profile obtained from AFM analysis (**Fig. 2** (c)) confirms larger Co grain sizes with a height up to 220 nm and a length up to 250 nm. The role of the boric acid during Mo deposition is therefore evident: in contrary to the Co deposition, the Mo deposition proceeds as a continuous film due to the boric acid blocking growth sites on the semiconductor surface and thereby significantly reducing the overall particle growth.

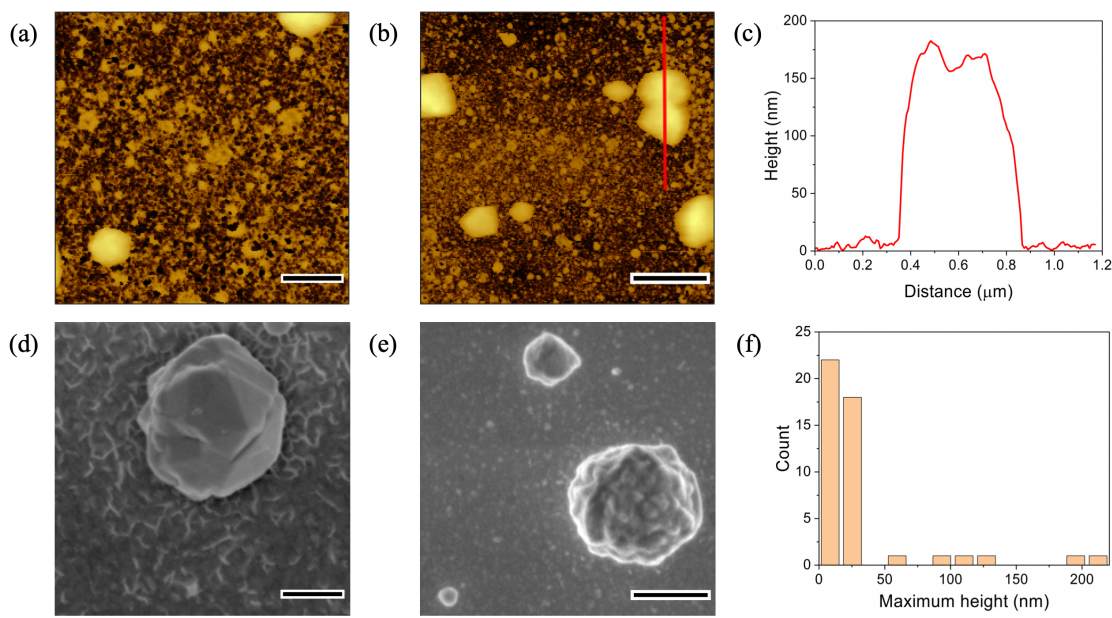

**Fig. 2** Optical analyses of the p-InP-Mo-Co photoelectrodes prepared via consecutive photoelectrodeposition from Mo solution I (10 mM Na$_2$MoO$_4$, 1 M Na$_2$SO$_4$, and 0.5 M H$_3$BO$_3$) and Co solution II (0.1 M CoCl$_2$, 0.5 M NaCl and 0.5% v/v 2-propanol). Mo deposition occurred at a potential of -0.09 V *vs* RHE applied for 10 s. Subsequent Co deposition occurred at the same potential applied for 3 s. An Xe arc lamp with a power density of 100 mW/cm$^2$ was used for illumination. AFM images of the electrode surface are shown with a scale bar of 200 nm (a) and 500 nm (b) with a corresponding particle height analysis (c) across 700 nm of the electrode surface. (d) and (e) show SEM images of the surface with scale bars of 200 nm and 500 nm, respectively. (f) provides an overview of the maximum

particle heights found on the electrode surface.

### 3.1.2 Consecutive photoelectrodepositions from Na$_2$MoO$_4$ - CoSO$_4$ electrolytes

The consecutive photoelectrodeposition using Mo solution I and Co solution I leads to relatively small Co depositions on the photoelectrode surface: EDX analysis indicates a Co:Mo ratio of 7:93 at % for the overall surface. The height of Co deposits ranges between 70 - 80 nm with a total particle length of about 250 nm (**Fig. 3** (a)-(f)), indicating again the particle growth-inhibiting role of the boric acid in the Co solution I. The Mo film shows a homogenous distribution of small Mo particles (≤ 20 nm), mostly present in the oxidation states Mo$^{4+}$ (75%, 230.6 eV/ 233.7 eV) and Mo$^{6+}$ (24.5%, 232.0 eV/ 235.1 eV). Mo$^0$ is present with 0.5 % (229.1 eV/ 232.2 eV, **Fig. S6** (b)). Co is only deposited as Co$^{2+}$ (**Fig. S6** (c)). EDX analysis shows a Co:Mo at% ratio within one particle of 80:20, and a Co:Mo ratio of 2:98 for the underlying Mo film. This demonstrates again that a large area of the electrode surface contains only one metal species and the desired proximity of Co and Mo on the surface is only partially present.

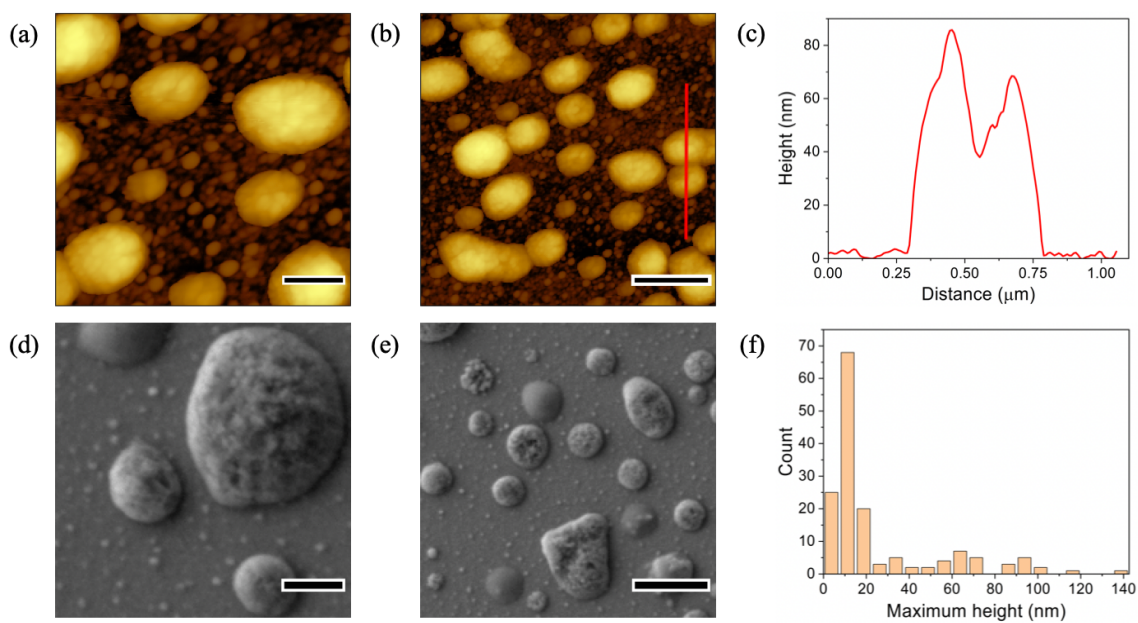

**Fig. 3** Optical analyses of the p-InP-Mo-Co photoelectrodes prepared via consecutive photoelectrodeposition from Mo solution I (10 mM Na$_2$MoO$_4$, 1 M Na$_2$SO$_4$, and 0.5 M H$_3$BO$_3$) and Co solution I (0.2 M CoSO$_4$, 1 M NaSO$_4$, 0.5 M H$_3$BO$_3$). Mo deposition occurred at a potential of -0.09 V *vs* RHE applied for 10 s. Subsequent Co deposition occurred at the same potential applied for 3 s. An Xe arc lamp with a power density of 100 mW/cm$^2$ was used for illumination. AFM images of the electrode surface are shown with a scale bar of 200 nm (a) and 500 nm (b) with a corresponding particle height analysis (c) across 1.05 μm of the electrode surface. (d) and (e) show SEM images of the surface with scale bars of 200 nm and 500 nm, respectively. (f) provides an overview of the maximum particle heights found on the electrode surface.

### 3.1.3 Consecutive photoelectrodepositions from MoCl$_3$ - CoSO$_4$ electrolytes

Using this electrolyte combination, EDX analysis shows a Co:Mo ratio of 25:75 on the electrode surface with a Co:Mo at % ratio of 68:32 within one particle and 34:66 for the underlying Mo film. The nearly equal height distribution is confirmed through AFM image analysis and shows an average particle height of 40 nm with most particles exhibiting maximum heights ≤ 60 nm (**Fig. 4** (a)-(f)). It has previously been reported that smoother deposition films are achievable with small additions of 2-propanol to the electrolyte[57], although the effect remains to be studied in further depth. As observed with the previous deposition solutions based on CoSO$_4$, metallic deposition of Co is not observed in XPS studies (**Fig. S7** (c)). Only Co$^{2+}$ is present on the surface, as binding energies of 781.3 eV/783.1 eV and 786.9 eV/791.3 eV suggest. Mo$^0$ (3%) was found at binding energies of 228.75 eV/231.90 eV, along with Mo$^{4+}$ (65.5%) at 230.8 eV/233.9 eV and Mo$^{6+}$ (31.5%) at 232.3 eV/235.5 eV.

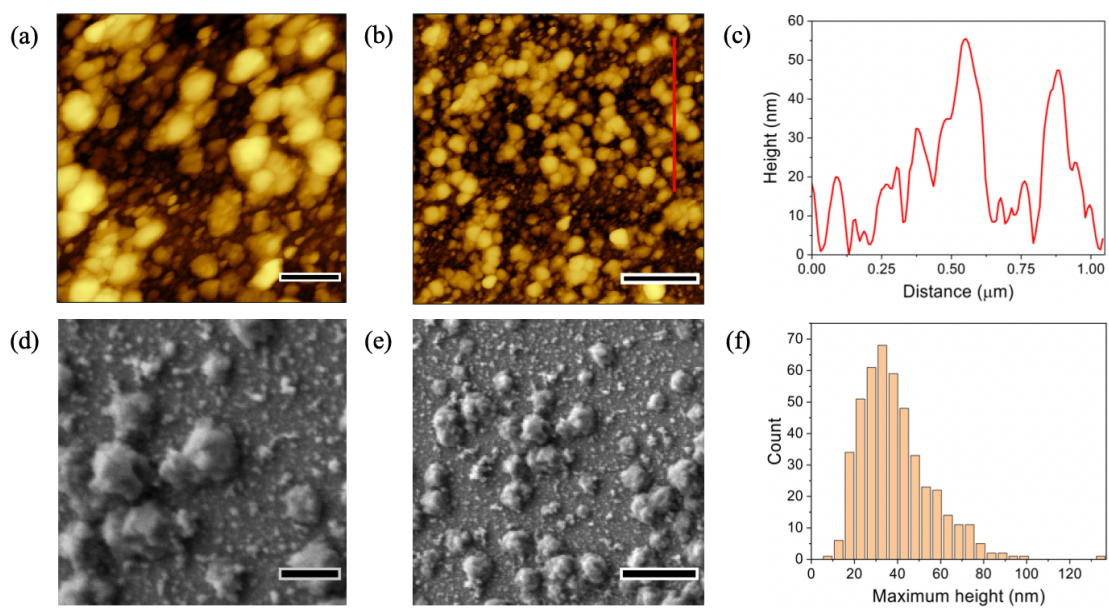

**Fig. 4** Optical analyses of the p-InP-Mo-Co photoelectrodes prepared via consecutive photoelectrodeposition from Mo solution II (1 mM MoCl$_3$, 0.5 M NaCl and 0.5% (v/v) 2-propanol) and Co solution I (0.2 M CoSO$_4$, 1 M Na$_2$SO$_4$ and 0.5 M H$_3$BO$_3$). Mo deposition occurred at a potential of -0.09 V *vs* RHE applied for 10 s. Subsequent Co deposition occurred at the same potential applied for 3 s. An Xe arc lamp with a power density of 100 mW/cm$^2$ was used for illumination. AFM images of the electrode surface are shown with a scale bar of 200 nm (a) and 500 nm (b) with a corresponding particle height analysis (c) across 1.05 μm of the electrode surface. (d) and (e) show SEM images of the surface with scale bars of 200 nm and 500 nm, respectively. (f) provides an overview of the maximum particle heights found on the electrode surface.

### 3.1.4 Consecutive photoelectrodepositions from MoCl$_3$ - CoCl$_2$ electrolytes

Using two chloride-based electrolytes for the Mo depositions results in a total Co:Mo ratio of 45:55 on the electrode surface and a smooth film with larger, granular deposits (**Fig. 5**). EDX analysis reveals that an individual grain consists of 95% Co and the film underneath is composed of 90% Mo. Whilst the surface initially shows equal amounts of Mo and Co, it is evident that both elements are still separated on the electrode surface unlike previously observed with depositions using $MoCl_3$- and $CoSO_4$-based electrolytes (**Fig. 4**). This deposition method results in 7.6% of metallic Co (**Fig. SI 8**), with a peak at 777.73 eV. $Mo^0$ is not observed. Instead, the Mo 3d spectrum shows $Mo^{4+}$ at 230.7 eV/233.8 eV and $Mo^{6+}$ at 232.2 eV/235.3 eV, respectively. In comparison, the $MoCl_3$(aq) electrolyte resulted in a metallic Mo deposition in the previous deposition using an electrolyte combination of $MoCl_3$(aq) and $CoSO_4$(aq).

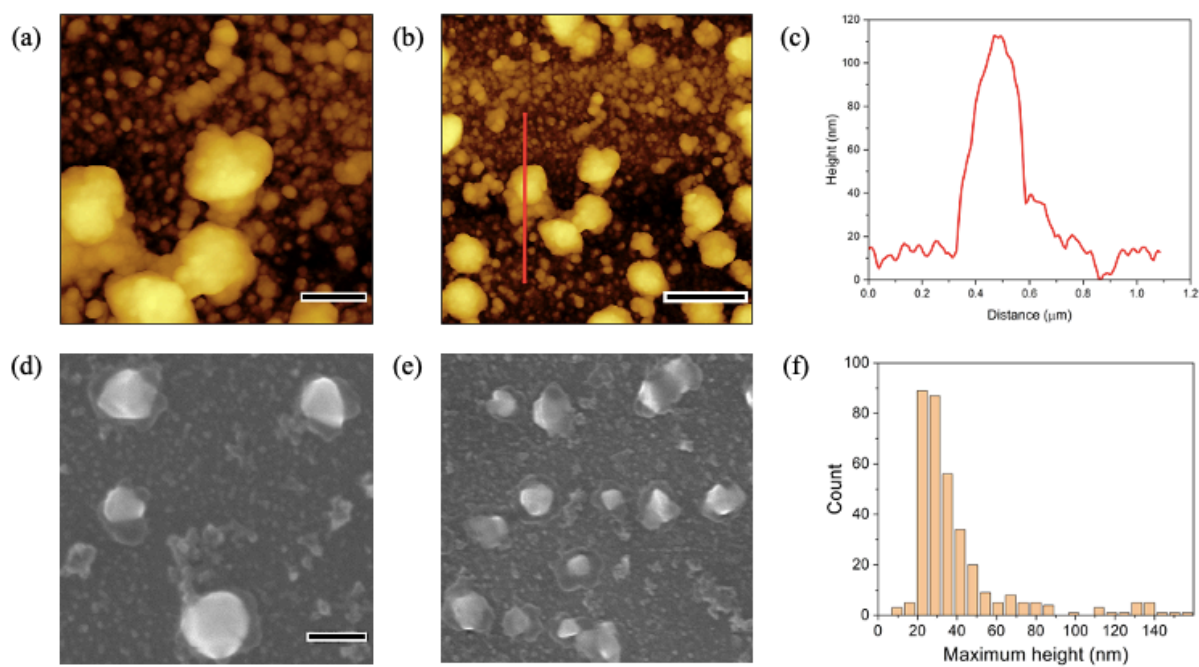

**Fig. 5** Optical analyses of the p-InP-Mo-Co photoelectrodes prepared via consecutive photoelectrodeposition from Mo solution II (1 mM $MoCl_3$, 0.5 M NaCl and 0.5% (v/v) 2-propanol) and Co solution II (0.1 M $CoCl_2$, 0.5 M NaCl and 0.5% v/v 2-propanol). Mo deposition occurred at a potential of -0.09 V *vs* RHE applied for 10 s. Subsequent Co deposition occurred at the same potential applied for 3 s. An Xe arc lamp with a power density of 100 mW/cm² was used for illumination. AFM images of the electrode surface are shown with a scale bar of 200 nm (a) and 500 nm (b) with a corresponding particle height analysis (c) across 1.1 μm of the electrode surface. (d) and (e) show SEM images of the surface with scale bars of 200 nm and 500 nm, respectively. (f) provides an overview of the maximum particle heights found on the electrode surface.

Concluding from the consecutive photoelectrodepositions, it is evident that the close arrangement of metallic Mo and Co on the photoelectrode is generally difficult to achieve, possibly due to the fast oxidation of the metals on the electrode surface when the electrode is removed from one deposition solution and placed into the subsequent one. Whereas the different Mo oxidation states achieved during the deposition might actually be of advantage for the NRR as highlighted by Yandulov and Schrock - who demonstrated that nitrogen reduction occurs at the Mo centre cycling between the Mo(III) and Mo(VI) oxidation states[50,58] - it is the metallic Co which lowers the nitrogen adsorption energy on the electrode surface and thus allows its protonation.[22]

## 3.2 Co-Mo bimetallic photoelectrodepositions

In addition to exploring the results of consecutive depositions, a protocol was developed for the co-photoelectrodeposition of Co and Mo from the same electrolyte in order to achieve $Mo^0$ and $Co^0$ depositions. The protocol for Co-Mo co-deposition was adapted from the works of Messaodi et al.[26], and the full details are provided in the Experimental Section. The deposition for the metals was carried out at -0.09 V *vs* RHE and resulted in the CA characteristics shown in **Fig. 6**. Depositions were explored in the presence and absence of boric acid to study the difference of the obtained surface morphology.

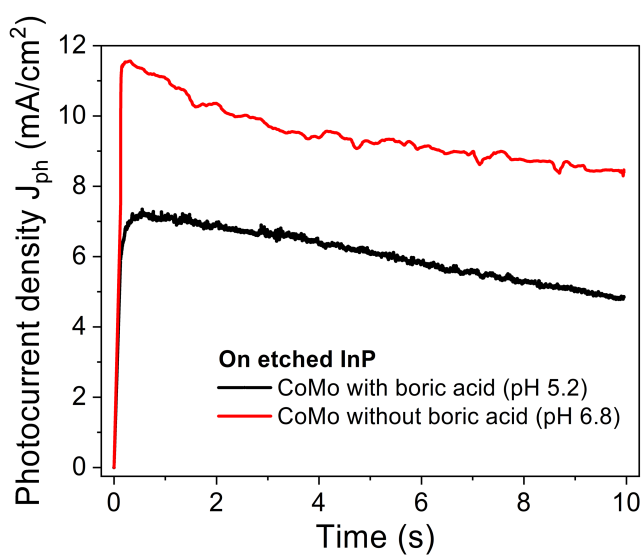

**Fig. 5** Chronoamperometric characteristics of co-photoelectrodepositions of Co and Mo in the presence and absence of 0.5 M boric acid in the electrolyte. A potential of -0.09 V *vs* RHE was applied for 10 s for both depositions. Illumination was provided by an Xe arc lamp at a power density of 100 mW/cm$^2$.

In acidic solutions containing positive metal ions, $MoO_4^{2-}$ forms tetrahedral $Mo_{12}O_{40}^{8-}$ complexes with a tetrahedral hole for the metal ion,[50] so-called heteropolymolybdates. These complexes are likely present in the deposition electrolytes used here, resulting in a close structural proximity of Mo and Co during deposition. The transient *J-t* characteristics show a clear difference in the presence of boric acid: despite that a lower solution pH (pH 5.2) initially predicts a higher hydrogen production rate, the overall photocurrent density is lowered by about 4 mA cm$^{-2}$. As discussed above, this is likely attributed to the inhibition of the particle growth rate, resulting in a large surface coverage with small particles. Particle height analyses accompanying the AFM and SEM images in **Fig. 7** confirm the hypothesis, indicating a homogeneous film with nanoparticles of about 30 nm in height. The cross-sectional TEM image in **Fig. S9** (a) indicates a uniform particle growth, leading to a cauliflower-shaped nanostructure. EDX analysis indicates an average ratio for Co:Mo of 60:40, despite that the initial molar Co:Mo ratio in the deposition electrolyte was 20:1. The phenomenon has also been observed in the literature, where it has been hypothesised that the Mo deposition process from $MoO_4^{2-}$ is favoured in the presence of boric acid and other metal ions in the solution.[30] This can likely be attributed to the heteropolymolybdate complexes that form and act as precursors for the deposition. In comparison, the Co:Mo ratio observed for depositions without boric acid is 90:10 (see below). XPS analysis of the surfaces was carried out to fully characterize the oxidation states of the metals present on the surface. In the presence of boric acid, evidence of metallic Mo is found as well as deposits of metal oxides with Mo in the oxidation states $Mo^0$, $Mo^{2+}$, $Mo^{4+}$ and $Mo^{6+}$. Cobalt is found in oxidation states ranging from $Co^0$ to $Co^{2+}$. The binding energy 777.87 eV is attributed to $Co^0$ and 227.73 eV /230.88 eV to $Mo^0$.[31] $Co^0$ makes up 30% of the overall peak, and $Mo^0$ about 15%. This demonstrates that co-photoelectrodeposition can indeed be employed to obtain deposits of metallic cobalt and molybdenum in close proximity on the electrode surface.

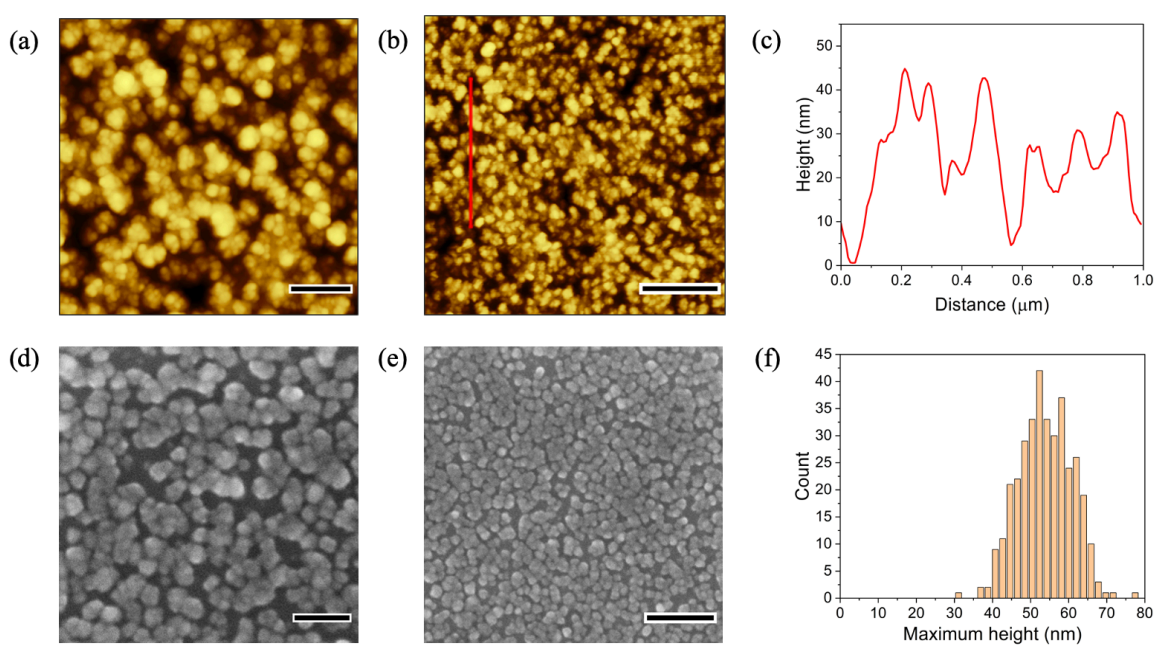

**Fig. 6** Optical analyses of the p-InP-CoMo photoelectrodes prepared via co- photoelectrodeposition from an aqueous solution of 0.2 M $CoSO_4$, 10 mM $Na_2MoO_4$, 1 M $Na_2SO_4$ and 0.5 M $H_3BO_3$ at a potential of -0.09 V *vs* RHE applied for 10 s. An Xe arc lamp with a power density of 100 mW/cm$^2$ was used for illumination. AFM images of the electrode surface are shown with a scale bar of 200 nm (a) and 500 nm (b) with a corresponding particle height analysis (c) across 1 μm of the electrode surface. (d) and (e) show SEM images of the surface with scale bars of 200 nm and 500 nm, respectively. (f) provides an overview of the maximum particle heights found on the electrode surface.

A much more incoherent surface morphology is obtained without boric acid in the deposition electrolyte, with regions of Co islands and branched networks of Co-Mo. The height distribution (**Fig. 8**) is observed to not be as symmetric. The average particle size ranges from 75 - 100 nm, but also larger particles with a size of about 300 nm are observed, indicating the irregularity of the deposited film. Interestingly, without boric acid, $Co^0$ and $Mo^0$ are not observed (**Fig. S10**). This demonstrates once more the importance of the electrolyte addition as a buffering agent, quenching hydroxide ions in proximity to the electrode surface which leads to metallic deposits of Co and Mo.

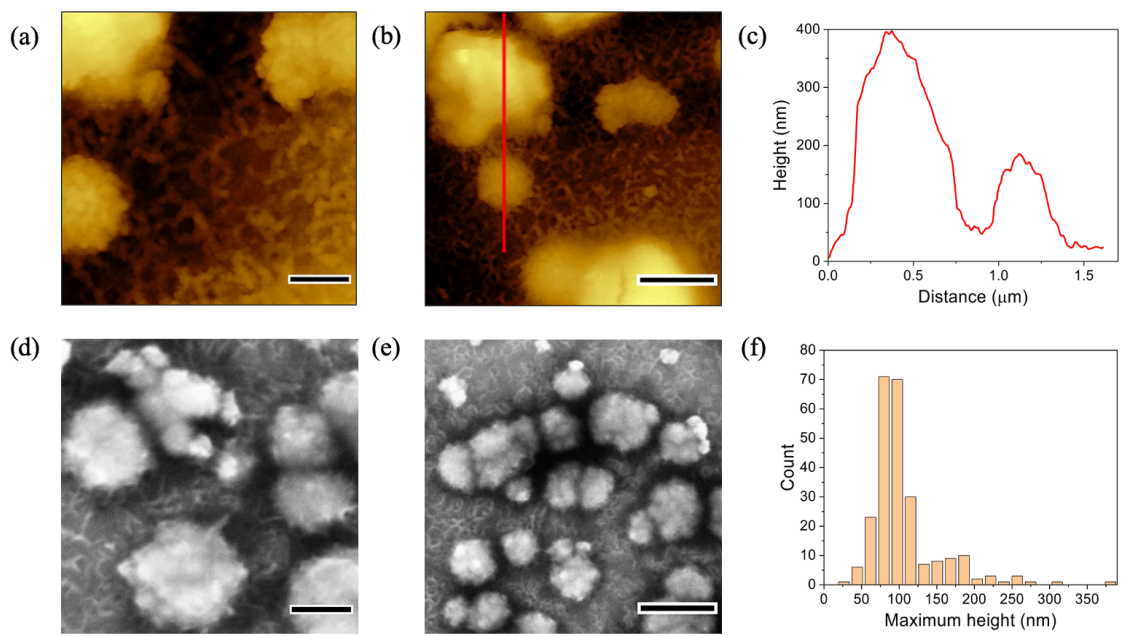

**Fig. 7** Optical analyses of the p-InP-CoMo photoelectrodes prepared via co- photoelectrodeposition from an aqueous solution of 0.2 M CoSO$_4$, 10 mM Na$_2$MoO$_4$ and 1 M Na$_2$SO$_4$ at a potential of -0.09 V *vs* RHE applied for 10 s. An Xe arc lamp with a power density of 100 mW/cm$^2$ was used for illumination. AFM images of the electrode surface are shown with a scale bar of 200 nm (a) and 500 nm (b) with a corresponding particle height analysis (c) across 1.6 μm of the electrode surface. (d) and (e) show SEM images of the surface with scale bars of 200 nm and 500 nm, respectively. (f) provides an overview of the maximum particle heights found on the electrode surface.

### 3.3 Co-photoelectrodepositions of Co, Mo and Ru

With the pursuit of exploring the synthesis of multi-metal electrocatalysts for NRR directly on the photoelectrode surface,[16,59,60] Ru photoelectrodepositions are explored onto the p-InP electrodes after Co-Mo deposition and alongside the deposition of Co and Mo (**Fig. 9**) as Ru represents another promising metal catalyst for the NRR.[61–63] As the further exploration of this synthesis strategy is part of current research efforts, only initial studies of surface compositions and morphologies are reported. Co-Mo deposits are firstly obtained via photoelectrodeposition from an aqueous solution of CoSO$_4$ (0.2 M), Na$_2$MoO$_4$ (10 mM), Na$_2$SO$_4$ (1 M), and H$_3$BO$_3$ (0.5 M) at a potential of -0.09 V *vs* RHE applied for 10 s. Subsequent Ru deposition occurs from an aqueous solution of RuCl$_3$ (0.33 mM), NaCl (0.5 M) and 2-propanol (0.5% v/v) at a potential of -0.09 V *vs* RHE applied for 5 s (**Fig. S11**). According to first EDX analyses, this p-InP-CoMo-Ru electrodes yields a Co:Mo:Ru ratio of 61:36:3, suggesting that the 5 s deposition of Ru successfully results in a low loading of Ru on the electrode. The height profile analysis (**Fig. S11** (c) and (f)) suggests that a majority of particles exhibits heights of about 60 - 100 nm. When Co, Mo and Ru are deposited from one electrolyte solution containing CoCl$_2$ (2 mM), MoCl$_3$ (0.1 mM), RuCl$_3$ (0.1 mM), H$_3$BO$_3$ (0.1 M), and NaCl (0.5 M) at a potential of -0.09 V *vs* RHE applied for 5 s, much smaller particles are observed, showing height profiles of 10 - 35 nm (**Fig. S12**). EDX analysis confirms the presence of Ru, showing a Co:Mo:Ru ratio of 8:82:10.

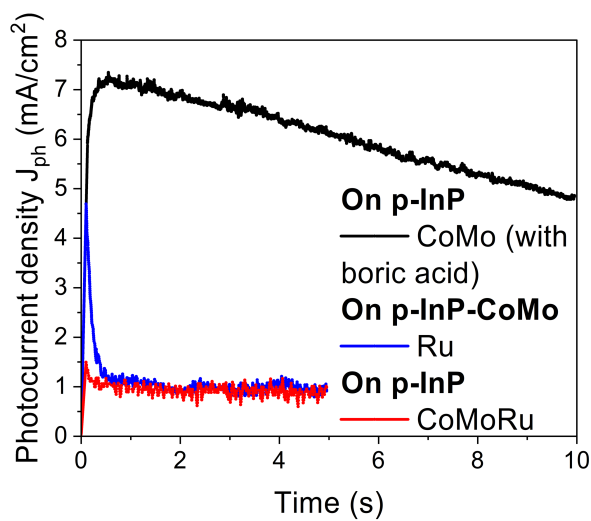

**Fig. 8** Chronoamperometric characteristics for the co-photoelectrodepositions of Co-Mo, followed by the deposition of Ru and the co-photoelectrodeposition of Co-Mo-Ru from one electrolyte solution containing $CoCl_2$ (2 mM), $MoCl_3$ (0.1 mM), $RuCl_3$ (0.1 mM), $H_3BO_3$ (0.1 M), and NaCl (0.5 M). All plots were recorded at a potential of -0.09 V *vs* RHE at the indicated durations. During the depositions, illumination occurred with an Xe arc lamp at a power density of 100 mW/cm² through a Quartz window.

Extensive studies have shown that coatings obtained by co-deposition of metals can show properties that are not obtained by the deposition of single metals due to alloy formation. As alloys and particularly high-entropy alloys have become increasingly popular also for the NRR, magnetometry is used to further elucidate the nature of Co and Mo deposits on the different photoelectrodes and conclude on the presence of metals or metal alloys as well as the general magnetic properties of the synthesised films.

### 3.4 SQUID Magnetometry

Magnetic spin is becoming an increasingly interesting parameter to explore and exploit in catalysis. Garcés-Pineda et al. describe the role of an external magnetic field in the oxygen evolution reaction (OER) under alkaline conditions where the magnetic field essentially works like a chiral centre, speeding up the dominant pathway to formation of the triplet state $O_2$ molecule, over the $H_2O_2$ side product.[64] Several other OER studies have reported on role of spin-coupled catalysis,[64–66] with literature available on the NRR as well.[67–70] Reported effects of spin-coupled catalysis on NRR include the enhanced adsorption and cleavage of $N_2$ molecules, the stabilisation of the rate-determining step of the reaction and the first protonation step to form *NNH. Current research efforts aim at finding better promoters for reducing the spin moment at the surface in order to lower the energy of the involved transition states and consequently, the energy barrier for the reaction.[67] Co-Mo alloys have been identified as soft-ferromagnetic materials[26,29,33,34] which can potentially also aid their catalytic performance. Here, we use this characteristic to investigate the potential formation of Co-Mo alloys during the consecutive and co-photoelectrodepositions.

The magnetisation (M) *vs* magnetic field (H) and temperature (T) plots for all four consecutively deposited p-InP-Mo-Co electrodes are given in the SI (**Fig. S13 - S16**). The signals are generally weak and made up of four contributions: a Pauli-like, temperature independent magnetic susceptibility from the metals in the films, a temperature dependent Curie-Weiss term due to isolated paramagnetic moments, a soft ferromagnetic component leading to a small amount of hysteresis in some of the *M-H* and *M-T* transients, and a significant diamagnetic signal, originating largely from the substrate which has been subtracted from the *M-H* plots shown in the SI. The ferromagnetic component persists up to 300 K, suggesting that the ordering temperature of this component is above room temperature. These 'mixed' magnetic properties have to however be considered with care due to the generally very weak signals which suggest that quantitative conclusions about the nature of the ordering and the anisotropy of the films is not practical. The similar behaviour observed in all four electrodes means that the magnetic response is unlikely due to any contamination issues and reflects the intrinsic magnetic properties of the consecutively deposited films.

Magnetisation *vs* field characteristics recorded at 5 K and 300 K for the co-deposited films prepared with and without boric acid are shown in **Fig. 10** and **11** (a) and (b), respectively. After estimating and subtracting the diamagnetic contribution from substrate, both samples reveal the typical s-shape behaviour with a low coercivity, indicative of a soft ferromagnetic material and attributable to the presence of a Co-Mo alloy. The remanent magnetisation, $M_r$, and coercive field, $H_c$, are higher in the film prepared without boric acid, indicating a harder ferromagnetic character. For both films, the magnetic response is largely isotropic and little changed for the field applied either parallel or perpendicular to the plane of the film. As expected, both $M_r$ and $H_c$ decrease at higher temperature. The magnetisation at high fields is larger for the film prepared without boric acid. In both cases, *M* does not fully saturate. This is probably due to a Pauli paramagnetic contribution to the signal from the metal film, but may in part be the result of the uncertainty in estimating the diamagnetic contribution from the film substrate.

The temperature dependence of the magnetisation for both films between 5 K and 300 K in an applied field of 250 Oe is also shown in **Fig. 10** and **11** (c) and (d). Hysteresis between the zero-field-cooled (ZFC) warming and field-cooled-cooling (FCC) curves confirms that the magnetic response contains a ferromagnetic component with an ordering temperature in excess of 300 K. An upturn in *M* at lower temperature is most likely due to a Curie-Weiss contribution from isolated paramagnetic moments.

Overall, the magnetic data for the consecutively deposited Co-Mo electrodes lead to inconclusive results, as indications of a strong paramagnetic character with a weakly ferromagnetic component are present. Both co-deposited films contain Co-Mo alloys with soft ferromagnetic properties as reported in the literature,[26,29,33,34] whereas films deposited with boric acid in the electrolyte yield a softer ferromagnet than those without boric acid.

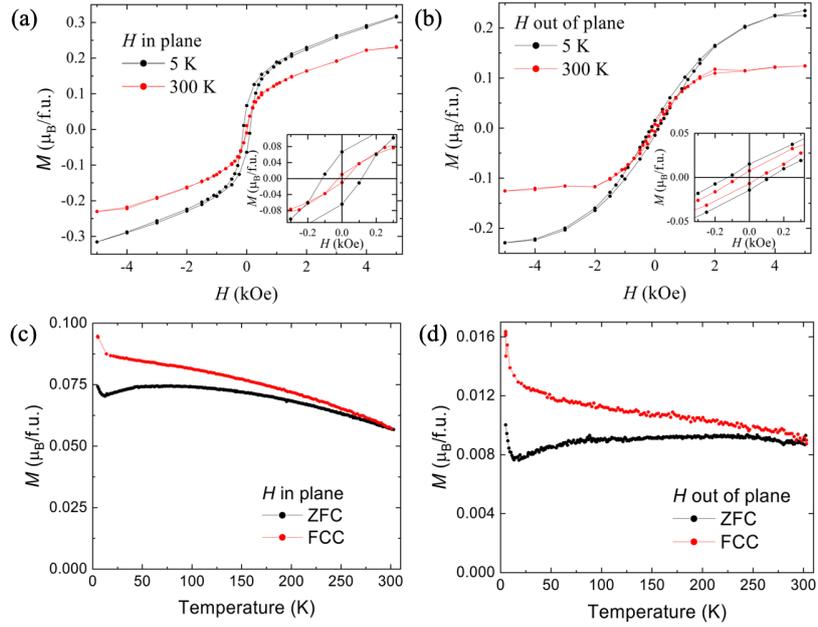

**Fig. 10** Magnetisation *vs* applied field plots at 5 K and 300 K with a magnetic field applied in plane (a) and out of plane (b) of the deposited film. Magnetisation *vs* temperature data were collected in a magnetic field of 250 Oe with the field applied either parallel (c) or perpendicular to the plane (d) of the film made via co-photoelectrodepositions of Co and Mo in the presence of boric acid.

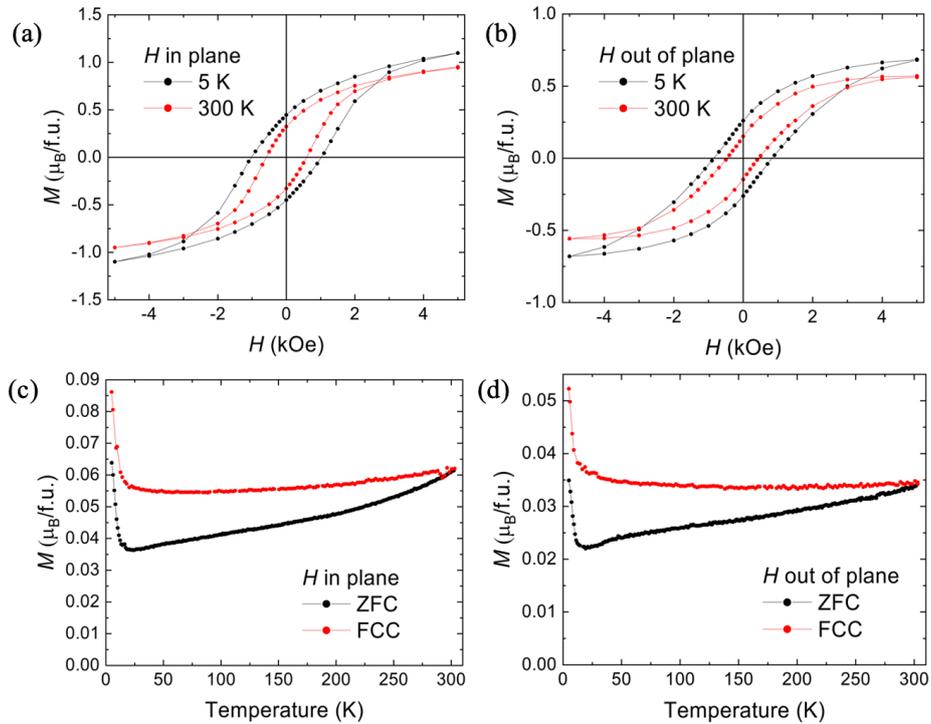

**Fig. 11** Magnetisation *vs* applied field plots at 5 K and 300 K with a magnetic field applied in plane (a) and out of plane (b) of the deposited film. Magnetisation *vs* temperature data were collected in a magnetic field of 250 Oe with the field applied either parallel (c) or perpendicular to the plane (d) of the film made via co-photoelectrodepositions of Co and Mo in the absence of boric acid.

## 3.5 Photoelectrochemical dinitrogen activation in aqueous electrolyte

Given the close proximity of $Co^0$ and $Mo^0$ in the films achieved through co-deposition in the presence of boric acid, the p-InP-CoMo photoelectrodes were further on tested for photoelectrochemical dinitrogen activation. **Fig. 12** shows the achieved photocurrent densities during initial chronoamperometric studies in the presence of Ar and $N_2(g)$ in 50 mM aqueous $H_2SO_4$ solutions at -0.09 *vs* RHE and 100 mW cm$^{-2}$ illumination provided through an Xe arc lamp.

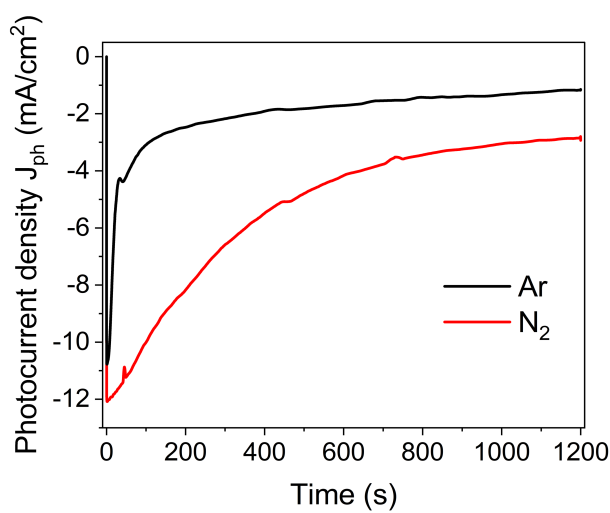

**Fig. 12** Chronoamperometric characteristics of photoelectrochemical $N_2(g)$ activation tests of the p-InP electrodes with a co-deposited Co-Mo electrocatalyst film in the presence of 0.5 M boric acid. Tests were carried out in 50 mM aqueous $H_2SO_4$ in the presence of Ar (black trace) and $N_2(g)$ (red trace) at -0.09 V *vs* RHE. Illumination was provided by an Xe arc lamp at a power density of 100 mW/cm$^2$.

The clear difference in the CA measurements in the presence of Ar and $N_2(g)$ provides first indication of an electrode reaction involving $N_2(g)$. Consecutive XPS studies of the electrodes tested in the presence of $N_2(g)$ reveal the presence of molybdenum-nitrogen interactions in the N 1s and Mo 3d spectra: the binding energy for Mo≡N is observed at 398.1 eV, for Mo-$NH_2$ at 400.3 eV and for Mo-$NH_3^+$ between 402.5 eV (**Fig. 13** (a)), which is in agreement with previously published Mo-N assignments[19]. The Mo-N (Mo (IV)) interactions are also observable in the Mo 3p$_{3/2}$ part of the spectrum at 397.8 eV. Similar evidence for reactions of molybdenum with nitrogen can be found in the Mo 3d spectrum, where all samples reveal significant changes in the Mo oxidation states after the photoelectrochemical test: although $Mo^0$ remains present on the surface, Mo (II), (IV) and (VI) are visible with clear evidence of $Mo^{6+}$-N species at 232.2 eV (Mo 3d$_{5/2}$) and 235.3 eV (Mo 3d$_{3/2}$), respectively (**Fig. 13** (b)). This provides a first indication that the fabricated p-InP-CoMo photoelectrodes are indeed capable of light-assisted dinitrogen reduction and consecutive protonation. Further, comprehensive photoelectrochemical tests of all electrodes also involving $^{15}N_2(g)$ as well as ionchromatographic tests for $NH_4^+(aq)$ formation are however required before final conclusions regarding the activity of all fabricated semiconductor - mixed-metal electrocatalyst systems can be made.

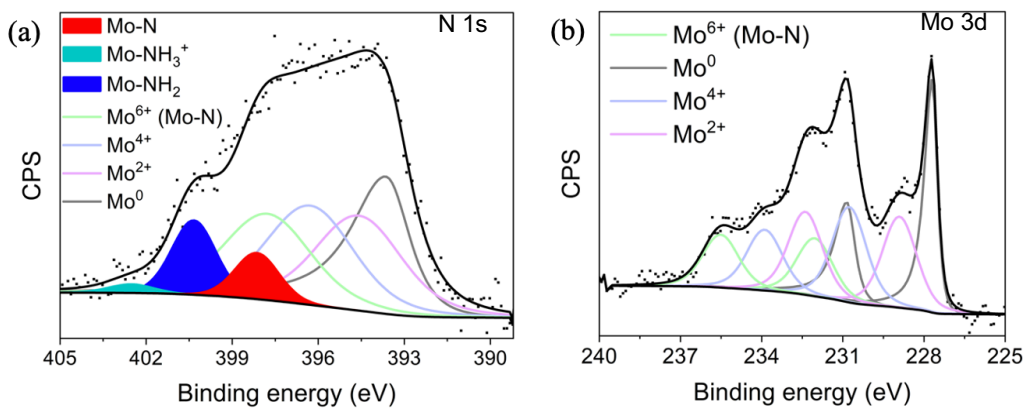

**Fig. 13** (a) N 1s XP spectrum overlapping with the Mo 3p$_{3/2}$ XP spectrum and (b) Mo 3d XP spectrum of the p-InP-CoMo electrodes prepared via co-photoelectrodeposition after application of -0.09 V *vs* RHE under constant N$_2$(g) purging in 50 mM H$_2$SO$_4$(aq) solution. An Xe arc lamp with a power density of 100 mW/cm$^2$ was used as an illumination source.

## 4. Conclusions

Photoelectrodeposition has been employed as a highly effective tool to fabricate Co, Mo and Ru mixed-metal electrocatalyst materials directly on the semiconductor surface. We have shown that Co-Mo alloys can be prepared on p-InP via co-photoelectrodeposition, whereas consecutive Co and Mo depositions lead to the presence of various metal oxides on the electrode surface with in-part large regioselectivities and a large variety of particle sizes. The synthesis of films containing Mo, Co and Ru was also explored by successfully depositing Ru on a p-InP-CoMo electrode via photoelectrodeposition and from an electrolyte containing the chloride salts of all three elements. The obtainable oxidation states and further properties of these films containing three elements are currently under further investigation.

Overall, our study demonstrates that a large variety of film compositions and characteristics are achievable with the different photoelectrodeposition techniques, electrolyte compositions and metal precursors explored here, leading to a precise control of the surface morphology, composition as well as properties. It outlines a pathway for a low-energy and cost-efficient synthesis of mixed metal electrocatalyst materials and alloys which could also be relevant e.g., within the framework of high-entropy alloy catalyst synthesis[16,59,60] and more generally, the fabrication of mixed-metal electrocatalysts for complex (photo-)electrocatalytic reactions such as the oxygen evolution reaction (OER) and carbon dioxide reduction reaction (CO$_2$RR),[71–73] requiring the transfer of multiple electrons and protons and the stabilisation of intermediate reaction steps. Initial photoelectrochemical tests with N$_2$(g) were furthermore carried out with the co-deposited CoMo photoelectrodes due to the presence of Co$^0$ and Mo$^0$ on the electrode surface. The chronoamperometric measurements showed a clear difference in the photocurrent densities observed in the presence of Ar and N2(g). XPS studies of the electrodes tested in the presence of N$_2$(g) reveal a significant variety of nitrogen-molybdenum interactions in the N 1s and Mo 3d spectrum after tests in 50 mM H$_2$SO$_4$(aq) which provides strong evidence for dinitrogen reduction and subsequent protonation. Further (photo-)electrochemical tests are however required to conclude on the actual applicability of the mixed-metal films synthesised here for the NRR, considering electrolyte optimizations, tests for $NH_{3(g)}/NH_4^+{}_{(aq)}$ formation as well as the utilization of $^{15}N_2(g)$ to fully verify the successful activation of dinitrogen.

## 5. Acknowledgements

MW acknowledges financial support from the EPSRC-funded Warwick Analytical Science Centre (EP/V007688/1). KB and MK would like to thank the Department of Chemistry, University of Warwick, for an awarded EPSRC studentship to MK. Both authors would also like to thank Thomas Barnes, Dr Ananthu Rajan and Dr. Sebastian Pike for helpful discussions.

## 6. Supporting Information

A supporting information is available free of charge from the RSC website or from the corresponding author.

## 7. Conflict of Interest

The authors declare no conflict of interest.

## 8. Data Availability Statement

The data supporting the findings of this study are available from the corresponding author upon reasonable request.